\documentclass[aps,pra,10pt,twocolumn,superscriptaddress, longbibliography]{revtex4-2}
\usepackage{amsmath, amssymb, amsthm, physics}
\usepackage[utf8]{inputenc}
\usepackage[countmax]{subfloat}
\usepackage{comment}
\usepackage{bbm}
\usepackage{graphicx}
\usepackage{lineno}
\usepackage{color}
\usepackage{hyperref}
\usepackage[normalem]{ulem}
\usepackage{epstopdf}
\usepackage{dsfont}
\usepackage{enumitem}
\usepackage[english]{babel}
\usepackage{graphicx}
\usepackage{float}
\usepackage{lipsum}
\usepackage{curve2e}
\usepackage{xurl}
\usepackage{hyperref}
\usepackage{xcolor}
\usepackage{cancel}
\usepackage{csquotes}
\usepackage{soul}
\usepackage{array}
\usepackage{float}

\usepackage{fancyhdr}

\usepackage{amsfonts}
\usepackage{amsbsy}
\usepackage{mathbbol}
\usepackage{mathtools}
\usepackage{wrapfig}
\usepackage{bbm}
\usepackage{braket}
\usepackage{bm}
\hypersetup{colorlinks=true, linkbordercolor=white, linkcolor=blue}
\usepackage{braket}
\newcommand{\Np}{{N_{\textrm ph}}}

\makeatletter
\DeclareRobustCommand{\bibfield}[2]{%
  \def\tempfield{#1}%
  \def\titlefield{title}%
  \ifx\tempfield\titlefield
    \textit{#2}
  \else
    #2
  \fi
}
\makeatother

\begin{abstract}
Models of interacting complex systems provide the fundamental statistical physics reference frame for the study and the understanding of associative memories, machine learning, and the dynamics of neural networks. On the other hand, simulating complex multi-synaptic interactions on a classical hardware is computationally demanding due to the super-linear scaling of the system complexity. Photonic quantum technologies provide a promising solution to these limitations by leveraging on their inherent speed and parallel processing ability in order to simulate complex networks. Recently, a connection between multiphoton processes and generalized {$p$}-body Hopfield models has been theoretically established. Here, we design and demonstrate an experimental platform that exploits single photons distributed across a set of optical modes, in which controlled arrays of binary phase shifters act as Ising-like neurons. We focus specifically on a fully connected Hopfield Hamiltonian with four-body local interaction terms, realized via two-photon processes. Through quantum simulations on programmable photonic processors, the study identifies three distinct regimes: a memory retrieval phase, a spin-glass memory ``black-out" phase, and a paramagnetic phase. Experimental results confirm successful memory retrieval at low storage capacities and temperatures, where the system consistently relaxes to fixed points with high memory overlap, effectively reconstructing the stored patterns. Future research will extend the platform design to investigate networks with local or dilute interactions, while advances in the realization of scalable photonic circuits will enable architectures that encompass very large numbers of interacting spins.
\end{abstract}

\begin{document}

\title{Observation of associative-memory retrieval and spin-glass phases on a photonic quantum simulator}

\author{Taira Giordani}
\affiliation{Dipartimento di Fisica, Sapienza Universit\`{a} di Roma, Piazzale Aldo Moro 5, I-00185 Roma, Italy}

\author{Gennaro Zanfardino}
\affiliation{Dipartimento di Medicina Sperimentale, Universit\`{a} del Salento, c/o Campus Ecotekne, Via Monteroni, 73100 Lecce, Italy}
\affiliation{Institute of Nanotechnology of the National Research Council of Italy, CNR-NANOTEC, Lecce Central Unit, c/o Campus Ecotekne, Via Monteroni, 73100 Lecce, Italy}
\affiliation{Dipartimento di Ingegneria Industriale, Universit\`{a} degli Studi di Salerno,
Via Giovanni Paolo II 132, 84084 Fisciano (SA), Italy}
\affiliation{INFN, Sezione di Napoli, Gruppo Collegato di Salerno, Italy}

\author{Luca Leuzzi}
\affiliation{Institute of Nanotechnology of the National Research Council of Italy, CNR-NANOTEC, Rome Unit, Piazzale A. Moro 5, I-00185, Rome, Italy }
\affiliation{Dipartimento di Fisica, Sapienza Universit\`{a} di Roma, Piazzale Aldo Moro 5, I-00185 Roma, Italy}

\author{Enrico Bonfissuto}
\affiliation{Dipartimento di Fisica, Sapienza Universit\`{a} di Roma, Piazzale Aldo Moro 5, I-00185 Roma, Italy}

\author{Eugenio Caruccio}
\affiliation{Dipartimento di Fisica, Sapienza Universit\`{a} di Roma, Piazzale Aldo Moro 5, I-00185 Roma, Italy}

\author{Gabriele Gasbarri}
\affiliation{Department of Computing Sciences, Bocconi University, Via Gugliemo R\"ontgen 1, I-20136 Milan, Italy}

\author{Mattia Bossi}
\affiliation{Istituto di Fotonica e Nanotecnologie, Consiglio Nazionale delle Ricerche (IFN-CNR), Piazza Leonardo da Vinci, 32, I-20133 Milano, Italy}
\affiliation{Dipartimento di Fisica - Politecnico di Milano, Piazza Leonardo da Vinci 32, 20133 Milano, Italy}

\author{Abhiram Rajan}
\affiliation{Istituto di Fotonica e Nanotecnologie, Consiglio Nazionale delle Ricerche (IFN-CNR), Piazza Leonardo da Vinci, 32, I-20133 Milano, Italy}
\affiliation{Dipartimento di Fisica - Politecnico di Milano, Piazza Leonardo da Vinci 32, 20133 Milano, Italy}

\author{Riccardo Albiero}
\affiliation{Istituto di Fotonica e Nanotecnologie, Consiglio Nazionale delle Ricerche (IFN-CNR), Piazza Leonardo da Vinci, 32, I-20133 Milano, Italy}

\author{Francesco Ceccarelli}
\affiliation{Istituto di Fotonica e Nanotecnologie, Consiglio Nazionale delle Ricerche (IFN-CNR), Piazza Leonardo da Vinci, 32, I-20133 Milano, Italy}

\author{Nicol\`o Spagnolo}
\affiliation{Dipartimento di Fisica, Sapienza Universit\`{a} di Roma, Piazzale Aldo Moro 5, I-00185 Roma, Italy}

\author{Raffaele Santagati}
\affiliation{Quantum Lab, Boehringer Ingelheim, 55218 Ingelheim am Rhein, Germany}

\author{Stefano Paesani}
\affiliation{NNF Quantum Computing Programme, Niels Bohr Institute,
University of Copenhagen, Blegdamsvej 17, DK-2100 Copenhagen Ø, Denmark}

\author{Marco Leonetti}
\affiliation{Institute of Nanotechnology of the National Research Council of Italy, CNR-NANOTEC, Rome Unit, Piazzale A. Moro 5, I-00185, Rome, Italy }
\affiliation{Center for Life Nano- \& Neuro-Science, Italian Institute of Technology, IIT, Rome, Italy}

\author{Roberto Osellame}
\email{roberto.osellame@cnr.it}
\affiliation{Istituto di Fotonica e Nanotecnologie, Consiglio Nazionale delle Ricerche (IFN-CNR), Piazza Leonardo da Vinci, 32, I-20133 Milano, Italy}

\author{Giorgio Parisi}
\affiliation{International Research Center of Complexity Sciences, Hangzhou International Innovation Institute, Beihang University, Hangzhou 311115, China}
\affiliation{Dipartimento di Fisica, Sapienza Universit\`{a} di Roma, Piazzale Aldo Moro 5, I-00185 Roma, Italy}
\affiliation{Institute of Nanotechnology of the National Research Council of Italy, CNR-NANOTEC, Rome Unit, Piazzale A. Moro 5, I-00185, Rome, Italy }

\author{Giancarlo Ruocco}
\email{giancarlo.ruocco@iit.it}
\affiliation{Center for Life Nano- \& Neuro-Science, Italian Institute of Technology, IIT, Rome, Italy}
\affiliation{Dipartimento di Fisica, Sapienza Universit\`{a} di Roma, Piazzale Aldo Moro 5, I-00185 Roma, Italy}

\author{Fabrizio Illuminati}
\email{fabrizio.illuminati@cnr.it}
\affiliation{Institute of Nanotechnology of the National Research Council of Italy, CNR-NANOTEC, Lecce Central Unit, c/o Campus Ecotekne, Via Monteroni, 73100 Lecce, Italy}
\affiliation{Dipartimento di Ingegneria Industriale, Universit\`{a} degli Studi di Salerno,
Via Giovanni Paolo II 132, 84084 Fisciano (SA), Italy}
\affiliation{INFN, Sezione di Napoli, Gruppo Collegato di Salerno, Italy}

\author{Fabio Sciarrino}
\email{fabio.sciarrino@uniroma1.it}

\affiliation{Dipartimento di Fisica, Sapienza Universit\`{a} di Roma, Piazzale Aldo Moro 5, I-00185 Roma, Italy}

\maketitle
\pagestyle{plain}

The modeling and understanding of Neural Networks (NNs) and their learning processes are fundamentally rooted in the study and simulation of complex systems. In this context, one of the most representative examples are Spin Glass (SG) models. SGs are disordered magnetic systems characterized by frustrated interactions, resulting in a rugged free-energy landscape with numerous stable and metastable states \cite{edwards1975theory, parisi1979infinite, parisi1983order, mezard1984nature}. This complexity, manifest in phenomena such as replica symmetry breaking, ultrametricity, and off-equilibrium aging dynamics, has established itself as a foundational paradigm for studying computational hardness and complex optimization in fields ranging from computer science to biology \cite{charbonneau2023spin}. 

In machine learning, SG systems provide a valuable framework for understanding the memory properties of NN architectures. This connection is best exemplified by the Hopfield model, a dense recurrent NN whose memory retrieval dynamics can be mathematically formulated as an SG system of statistical mechanics \cite{hopfield1982neural, amit1985storing,krotov2016dense}. In these architectures, specifically tailored synaptic connections enable the embedding of hidden memory patterns as dynamic attractors. This mechanism enables the network to map incoming inputs into learned states, thereby facilitating complex sequential data processing tasks such as text, speech, or time-series analysis \cite{ramsauer2021, krotov2023new}. 

Unfortunately, the computational complexity of simulating SGs or training Hopfield NNs scales superlinearly, often exponentially, with the system size. This severe limitation results in significant energy demands for modeling large-scale systems or high-capacity networks, creating a fundamental bottleneck both in fundamental research and in concrete applications. 

To address this challenge, hardware-level photonic solutions have been proposed and tested to simulate spin systems such as Ising machines \cite{McMahon2016, inagaki2016large, pierangeli2019large, wu2025monolithically}, and SG models \cite{pierangeli2021scalable, leonetti2021optical}. Photonic computing offers significant advantages over electronics, including high-speed operation, minimal interconnect crosstalk, and reduced energy dissipation per operation \cite{wetzstein2020inference, shastri2021photonics, mcmahon2023physics, liao2023integrated}. In these photonic architectures, spin state information can be physically encoded as relative phases distributed across distinct spatial optical modes. This configuration effectively defines a programmable transfer matrix that governs the optical interference and executes the desired analog computations. This control can be achieved using tunable phase shifters, which are a prominent feature of integrated photonic circuits \cite{wang2020integrated, pelucchi2022potential}. Such integrated architectures are particularly advantageous for optical computing, as they offer precise, fully controllable transfer matrices while ensuring inherent phase stability and robust scalability in the mapping of physical models \cite{Clements2016, carolan2015universal, pentangelo2024high}.

Previous studies have pioneered large-scale optical simulations using classical states of light and scattering processes. In such systems, the relationship between the input and output optical modes is described by a complex-valued transmission matrix that encodes the attenuation and phase shift for each pathway \cite{leonetti2021optical}. 
In this scenario, optical platforms offer a computational advantage for simulating fully connected systems: in conventional computing, evaluating the energy contributions from dense spin couplings is an explicit and computationally costly process. Conversely, in the optical paradigm, once a spin configuration is encoded, wave interference inherently computes all interaction terms simultaneously and at the speed of light, as they are instantaneously summed within the measured optical intensity \cite{chabaud2021quantum,mcmahon2023physics, pierangeli2021scalable, leonetti2021optical}. 
 
Moving from the classical to the quantum domain, the above paradigm can be broadly generalized by exploiting multiphoton nonclassical states of light \cite{Dellanno2006}, thus establishing a possible direct connection with photonic quantum computing, a field that has experienced tremendous growth in recent years, largely driven by the rapid development and increasing scalability of integrated photonic platforms \cite{flamini2019photonic, wang2020integrated, pelucchi2022potential,sparrow2018simulating, maring2024versatile, Hoch2025, aghaeerad2025scaling}. Indeed, it has been recently shown that the counting statistics of individual photon states within a multiport interferometer makes it possible to reproduce the dynamics of multi-synaptic dense NNs, and that the dynamics of photon pairs can simulate generalized $p$-body Hopfield models \cite{zanfardino2025}. Moreover, the evolution of ensembles of $n_{\rm ph}$ photons enables the direct simulation of dense architectures featuring interactions across $2n_{\rm ph}$ bodies. This approach ultimately unlocks a distinct digital photonic advantage that scales intrinsically with the complexity of the system, laying the foundations for the realization of quantum photonic spin simulators \cite{zanfardino2025}.

This work demonstrates the first experimental realization of a photonic Hopfield neural network using programmable integrated quantum photonics, in which two-photon quantum interference in reconfigurable universal interferometers emulates spin-glass dynamics and drives Metropolis Monte Carlo sampling to map out memory-retrieval phase diagrams. In particular, we investigate the various phases of disordered spin systems, namely the paramagnetic, spin glass, and memory retrieval phases of a generalized Hopfield model mapped onto 6- and 10-spin architectures. This is achieved using a quantum photonic platform known as QOLOSSUS \cite{Hoch2025,rodari2024,monbroussou2025photonic}, which is based on 8- and 12-mode laser-written universal integrated photonic circuits \cite{corrielli2021femtosecond,pentangelo2024high,monbroussou2025photonic}, to process path-encoded photonic quantum states. The system manipulates two photons emitted by a quantum-dot single-photon source and detects them with superconducting single-photon detectors. 

The extensive programming flexibility of this apparatus allows us to execute Metropolis Monte Carlo (MC) optimization \cite{metropolis1953equation} directly driven by the photonic simulator and enables a rigorous exploration of each thermodynamic phase. Furthermore, the high degree of control provides the capacity to dynamically manipulate the structure of the spin connections and to compute ensemble averages over different instances of random spin couplings. This capability experimentally confirms successful memory retrieval and provides a robust platform for simulating complex multi-synaptic networks. Ultimately, this work paves the way toward quantum neural networks and opens new frontiers by leveraging the computational advantage of photonic processing alongside quantum technologies \cite{steinbrecher2019quantum,chabaud2021quantum, monbroussou2025photonic}.

\begin{figure}[t!]
    \centering
    \includegraphics[width=0.99\columnwidth]{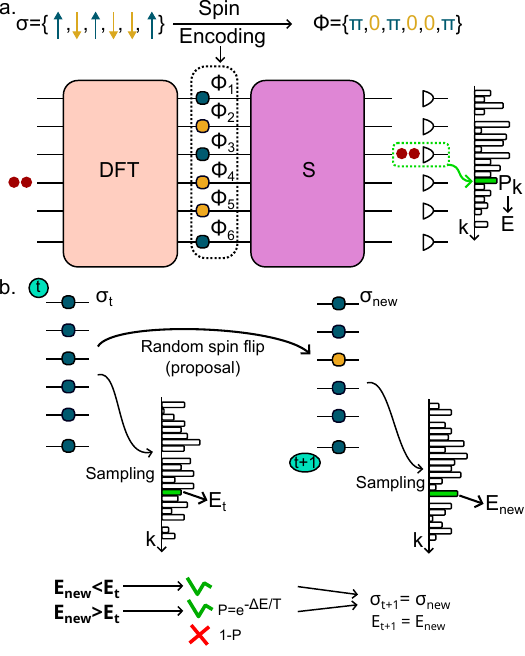}
    \caption{\textbf{Photonic quantum simulator of generalized Hopfield models}. 
    a) The two photons first propagate through a linear optical transformation represented by a Discrete Fourier Transform (DFT). Binary spin configurations are mapped onto discrete phase $\{\uparrow, \downarrow\}\rightarrow\{0, \pi\}$ values to physically modulate the optical field. Then, the optical modes are mixed by a final linear optical transformation $S$. 
    b) Metropolis optimization via dynamic sampling processes for energy evaluation: at a generic time step, the algorithm generates a proposal via a random spin flip to create a new physical configuration. The photonic spin simulator then directly computes the new associated energy. A classical routine evaluates the transition by comparing the newly computed energy against the previous one. Spin states updates strictly follow a physical acceptance criterion. The system accepts lower energy states deterministically and accepts higher energy states with probability 
    $P=e^{-\Delta E/T}$, where $\Delta E$ is the energy gap between the current best spin configuration and the one proposed by the spin flip, and $T$ is the temperature of the system. An accepted proposal successfully dictates the spin state for the subsequent temporal step in an equilibrium dynamics.}
    \label{fig:concept}
\end{figure} 

\begin{figure*}[t!]
    \centering    \includegraphics[width=1\linewidth]{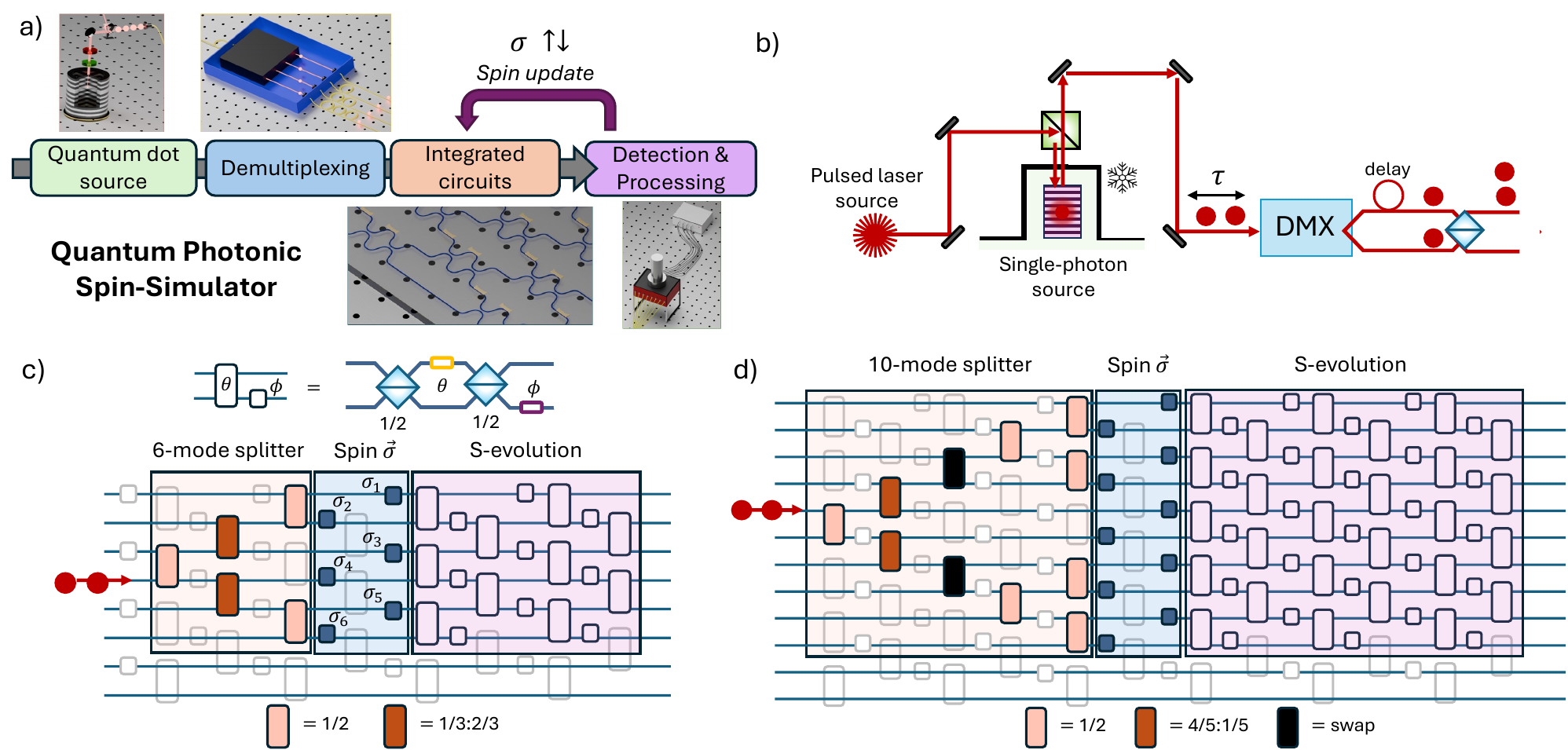}
    \caption{\textbf{Quantum photonic spin-simulator.} a) Overview of QOLOSSUS's overall apparatus. This setup includes a single photon source based on quantum dot technology and a time-to-space demultiplexing system (DMX) to synchronize two-photon states. The apparatus simulates spin dynamics within fully programmable integrated circuits. It also includes a detection and data processing stage to estimate the system energy and propose a new spin update according to the Metropolis algorithm. b) Details of the apparatus for the generation of the two-photon states. This stage comprises a pulsed laser to excite the quantum dot single photon source. The DMX system is followed by time delays to synchronize photons into different spatial modes. A symmetric beam splitter then performs the postselected generation of a two-photon input in a single spatial mode via Hong-Ou-Mandel interference. c) Layout of the 8-mode universal reprogrammable interferometer. The section highlighted in brown uniformly splits the single photon signal entering from input port 4 across 6 modes. Transparent rectangles indicate tunable beam splitters (Mach-Zehnder interferometers (MZIs) in the inset) configured with a transmissivity equal to 1. Transparent squares indicate phases set to 0. The legend reports the splitting ratios of the colored beam splitters. The blue region of the circuit encodes the 6 spins using six phase shifters capable of alternating between 0 and $\pi$. Finally, the purple section encodes the scattering evolution matrix $S$. This section comprises 10 variable beam splitters and 7 phases programmed to encode a specific scattering matrix. d) Analogous scheme generalized to encode a 10-spin system within the 12-mode chip. The expanded circuit depth facilitates scaling up the size of the system and provides higher reconfigurability for the $S$ matrix.}
    \label{fig:platform}
\end{figure*}

\begin{figure*}[t!]
    \centering    \includegraphics[width=1\linewidth]{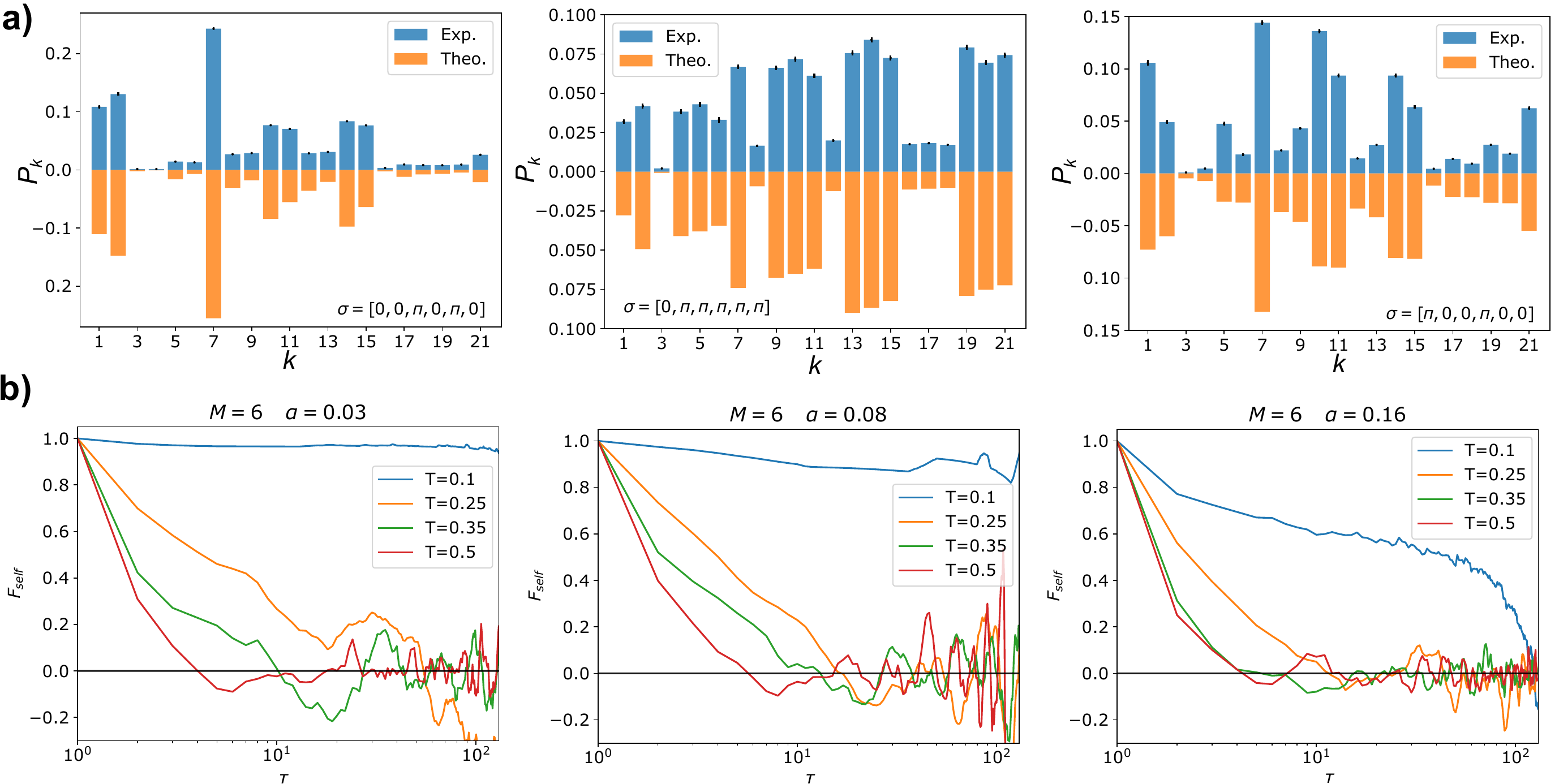}
    \caption{\textbf{Experimental characterization of the 6-spin simulator within the 8-mode integrated device.} a) Two-photon output distribution for three different spin configurations, which are indicated in the plot, for a fixed $S$ matrix. In blue, the expected distribution $P^{\mathrm{theo}}$; in orange, the experimental one $P^{\mathrm{exp}}$. The error bars in black derive from the Poissonian statistics of single-photon counts. The accordance with the theoretical distributions is quantified by the total variation distance (TVD) defined as $\frac{1}{2}\sum_k|P^{\mathrm{exp}}_k-P^{\mathrm{theo}}_k|$, where the index $k$ runs over the 21 (6 bunching + 15 no-bunching states) two-photon output configurations. The TVDs of the 3 spin settings are $0.069\pm 0.003,\, 0.054\pm 0.003, 0.136 \pm 0.003$. The average TVD over the 64 spin configurations is $0.1183 \pm  0.0004 $. b) Spin-spin autocorrelation function $F_{\rm self}(\tau)=\langle \bm\sigma(t)\cdot\bm\sigma(t+\tau)\rangle$ for various temperature of the MC simulation. One time step $\tau$ corresponds to $M=6$ spin updates proposals investigated in the optimizations. The plots show various runs at different temperatures, by increasing the parameter $\alpha=N_P/6^2$. The scattering matrix $S$ is the same as in panel a) for the two-photon output distribution.}
    \label{fig:8mode_results}
\end{figure*}

\section*{Photonic quantum simulation of complex systems: the concept}

The {quantum} photonic spin simulator is composed of three elements (see Fig.~\ref{fig:concept}\textcolor{blue}{a}): (i) an interferometer creating an input superposition state of {$n_{\mathrm{ph}}$} 
photons in $M$ modes; (ii) a layer of $M$ linear phase shifters, one per mode, used to encode a binary pattern into the photon state; (iii) an interferometric scattering matrix that determines the output state whose probability distribution {maps to the configuration energies in} {a $p$-body} Hopfield-like Hamiltonian {of $M$ neurons, with $p=2n_{\mathrm{ph}}$.}

The photonic quantum simulator is represented by means of Fock basis states $\vert \vec{x} \rangle$ of $n_{\mathrm{ph}}$ single photons in $M$ modes. Here, the vector $\vec{x} = (x_1,..., x_{n_{\mathrm{ph}}})$ represents the Fock state configuration:  
\begin{eqnarray}
    \ket{\vec{x}}=\prod_{i=1}^{n_{\mathrm{ph}}}\hat{a}_{x_{i}}^{\dagger} \ket{\bm 0} \, ,
\end{eqnarray}
where {$x_i \in \{1,\dots,M\}$} is the mode of the i-th photon,  $\ket{\bm 0} = \ket{0}^{\otimes M}$ {is the vacuum state over $M$ modes}.
Next, we associate to each mode $x_{i}$ a neuronal activity $\sigma_{x_{i}}=e^{i\phi_{x_{i}}}$, where $\phi_{x_{i}}$ is the phase shift imprinted to the $i$-th photon by the phase shifter in the mode $x_i$. 
The phases take the values
$\phi_{x_{i}} = 0$ or $\phi_{x_{i}} = \pi$ so that the phase shifts $\sigma_{x_{i}} = \pm 1$ are Ising variables. 
The spins encode a binary {configuration} of length $M$ into the photon state. In the following, whenever we refer to a {configuration}, we mean a given sequence of binary phases in the layer of controllable shifters. 

To realize the mapping, we consider the most general input state of {$n_{\mathrm{ph}}$} indistinguishable photons in $M$ modes: $\ket{\psi_0} = \sum_{\vec{x}\in \mathcal{C}} a_{\vec{x}} \vert \vec{x} \rangle \ $
where $\mathcal{C}$ is the set of all possible configurations of {$n_{\mathrm{ph}}$} 
single photons over $M$ modes, {whose cardinality is} {$|\mathcal{C}|= \binom{M+n_{\mathrm{ph}}-1}{n_{\mathrm{ph}}}$}, and $a_{\vec{x}}$ is the amplitude of the configuration $\ket{\vec{x}}$.
This input state undergoes a transformation through the layer of $M$ controllable linear phase shifters representing the binary {configuration}. This turns the input state into:

\begin{equation}
    \vert \psi_{{\sigma}} \rangle = \sum_{\vec{x}\in \mathcal{C}} a_{\vec{x}}\, e^{i\sum_{j=1}^{n_{\mathrm ph}}\phi_{x_j}}\ket{\vec{x}} = \sum_{\vec{x}\in \mathcal{C}} a_{\vec{x}} \prod_{j=1}^{n_{\mathrm{ph}}} \sigma_{x_j}\ket{ \vec{x}}  \, .
\end{equation}
        
We then let the state evolve through a linear interferometric network represented by the {$M\times M$ scattering} matrix $S$. The scattering amplitude for a {bosonic state} in the {output configuration $\vec{k} = (k_1,...,k_{n_{\mathrm{ph}}})$} can be written in terms of the {$S$ sub-matrix permanents as  \cite{scheel2004permanents}}:
 \begin{equation} \label{amplitude}
     \langle \vec{k} \vert \psi_{{S,}\sigma} \rangle =\sum_{\vec{x} \in \mathcal{C}}  a_{\vec{x}}\frac{\mathrm{Perm}(S_{\vec{k}|\vec {x}})}{\sqrt{\mu(\vec{x})\mu(\vec{k})}}\prod_i^{n_{\mathrm{ph}}}\sigma_{x_i} \, ,
 \end{equation}
where $\vert \psi_{S,\sigma} \rangle = S \ket{\psi_{\sigma}}$ is the state evolved by the action of the scattering matrix $S$; $\mu(\vec{w}) = \prod_{j=1}^{M}n_{w_{j}}$ is the multiplicity of the state $\ket{\vec w}$;
$n_{w_{j}}$ is the occupation number of the mode $w_{j}$, with $\sum_{j=1}^{M} n_{w_{j}}=n_{\mathrm{ph}}$; and the objects $S_{\vec k|\vec x}$ are the $n_{\rm ph}\times n_{\rm ph}$ sub-matrices of the scattering matrix $S$ with entries $(S_{\vec k|\vec x})_{ij} = S_{k_i, x_j}$, and are defined in Eq.~\eqref{scattering} below. Finally, the objects ${\rm Perm}(\cdot)$ are the matrix permanents.
The output probability distribution $P(\vec{k})$, namely the probability of finding the {$n_{\mathrm{ph}}$} photons in the output configuration $\vec{k}$, is given by the squared modulus $\vert \langle \vec{k} \vert \psi_{{S,}\sigma} \rangle \vert^2$ of the expression in Eq.~\eqref{amplitude}. 
Among the set of all possible output configurations $\mathcal{C}$ we select a subset of ``target" configurations $\mathcal{K} \subseteq \mathcal{C} $. The probability of finding the photons in any configuration belonging to $\mathcal{K}$ is then given by a summation over the target configurations:

\begin{equation}
\begin{split}
\label{mapping}
    {P_{S}(\bm \sigma,\mathcal K)} &=\sum_{\vec k\in \mathcal{K}}|\braket{\vec{k}|\psi_{S,\sigma}}|^2 = \sum_{\vec{x}, \vec{y} \in \mathcal{C}} J_{\mathcal{K}}(\vec{x}, \vec{y}) \prod_i^{n_{\mathrm{ph}}} \sigma_{x_i}\sigma_{y_i}\, ,
\end{split}
\end{equation}
with

{\begin{eqnarray}
\label{J}
    J_{\mathcal{K}}(\vec{x}, \vec{y}) &\equiv& 
    \sum_{\vec k \in \mathcal K} \mathbb X_{\vec x}^{(\vec k)}\mathbb X_{\vec y}^{(\vec k)*},  \end{eqnarray}}
    {and
    \begin{eqnarray}
      \label{X}
   \mathbb X_{\vec x}^{(\vec k)} &\equiv&  \frac{a_{\vec{x}}}{\sqrt{\mu(\vec{x})}}  
 \frac{\mathrm{Perm}(S_{\vec{k}|\vec {x}})}{\sqrt{\mu(\vec{k})}}    \, ,
\end{eqnarray}
where the $n_{\mathrm{ph}} \times n_{\mathrm{ph}}$ scattering sub-matrices $S_{\vec{k}|\vec{x}}$ are defined according to the following structure:
\begin{equation}
\label{scattering}
S_{\vec{k}|\vec{x}} =
\left(\begin{array}{cccc}
S_{x_1,k_1} & S_{x_1,k_2} &   \ldots   & S_{x_1,k_{\Np}} \\
S_{x_2,k_1} & S_{x_2,k_2} &   \ldots   & S_{x_2,k_{\Np}}  \\
\vdots        & \vdots        &            &  \vdots  \\
S_{x_{\Np},k_1} & S_{x_{\Np},k_2} &\ldots & S_{x_{\Np},k_{\Np}}
\end{array}\right) \quad ,
\end{equation}
and the functions $\mathrm{Perm}\left(S_{\vec{k}|\vec{x}}\right)$ denote their permanents.

Looking at Eq.~\eqref{mapping} we recognize that the probability {$P_{S}(\sigma,\mathcal K)$ realizes a}
Hopfield-like Hamiltonian of the form
\begin{equation}
\label{p_Hamiltonian}
    H_S(\bm \sigma|\mathcal{K}) = -M P_{S}(\bm \sigma,\mathcal{K})\, ,
\end{equation}
where the minus sign ensures that maximizing the probability minimizes the energy. This quantity is used in the Metropolis algorithm to accept a spin-flip proposal during the simulation of the spin dynamics (see Fig.~\ref{fig:concept}\textcolor{blue}{b}).

Indeed, the Hamiltonian for a $p$-spin multi-synaptic Hopfield model \cite{Baldi1987, Gardner1987, Abbott1988, Horn1988} is {of the form}
$H_{\xi}(\bm \sigma) = - \sum_{i_i...i_p}^N J_{i_1...i_p} \prod^p_l \sigma_{i_l}  $, {with $J_{  \vec i}\equiv \sum_{\mu} \prod_{l=1}^p \xi_{i_l}^{(\mu)}$} {and $\vec{i}=(i_{1},\dots,i_{p})$}, 
which {is of the same form as the r.h.s. in Eq.~\eqref{mapping}}.
We see that the number of synaptic connections, Eq.~\eqref{J}, between neurons in the photonic network is then {$p=2n_{\mathrm{ph}}$}.
The mapping works as follows: the number of patterns, Eq.~\eqref{X}, to be memorized by the system is the size $N_P=|\mathcal K|$ of the subset $\mathcal{K}$ of the output configurations.
The energy, Eq.~\eqref{p_Hamiltonian}, of configuration $\bm \sigma$
depends on the set  of selected memories $\mathbb X^{(\vec k)}$, $\vec k \in \mathcal K$. 
The number $N_P$ of memories planted into the neural network, Eqs.~(\ref{mapping})-(\ref{X}) cannot exceed {$|\mathcal{C}|$} and if $\mathcal{K} = \mathcal{C}$,  we get the trivial case {$P_{S}(\bm \sigma,\mathcal{K}) = 1$} $\forall \sigma $.

It is important to observe that the structure of the $p$-neuron interaction $J_{\mathcal{K}}(\vec{x}, \vec{y})$ appearing in Eq.~\eqref{J} differs from {the one} of a classical {$p$}-spin Hopfield memory model. 
Although Eq.~\eqref{J} still has the form of an additive Hebb's rule, the elements of a memory pattern,  
$\mathbb X^{(\vec k)}_{\vec x}$, see Eq.~\eqref{X},
are now different from the neuronal activities  $\sigma_x$:
(i) rather than a vector, as in the original Hopfield models \cite{hopfield1982neural,Baldi1987,Gardner1987}, now the memory pattern is a $n_{\mathrm{ph}}$-tensor; (ii) the elements of the tensor can take complex continuous values and not only $\pm 1$.
For this reason, although we are able to control the scattering matrix $S$, it is not always immediate to identify the optimal spin configuration $\bm \sigma$ corresponding to a given stored pattern, as we will see when illustrating the system phase diagram.

\section*{The photonic quantum simulator}
An overview of the experimental platform is shown in Fig.~\ref{fig:platform}\textcolor{blue}{a}. The two-photon input states are generated via a quantum-dot source, operating in a cryogenic environment, that is excited by a laser source and emits a train of single photons separated by a fixed time interval $\tau$, equal to the inverse of the laser source repetition rate. A time-to-spatial de-multiplexing system is then used to separate the single photons in two spatial modes. The photons are then synchronized via suitably prepared fiber delays, and made indistinguishable in the polarization degree of freedom. The prepared two-photon state is then a sequence of two-photon states impinging in the two input ports of a symmetric beam-splitter, where Hong-Ou-Mandel interference occurs \cite{HOM,Bouchard_2021}. By selecting one of the output ports of the beam-splitter, and post-selecting on the measurement of two-photon coincidences, this corresponds to generating a two-photon state in a single spatial mode (see Fig.~\ref{fig:platform}\textcolor{blue}{b}). 

The generated two-photon state is then injected in a 8-mode and 12-mode programmable interferometer, allowing for universal operation \cite{Hoch2025,monbroussou2025photonic}. The layout for the 8-mode device is shown in Fig.~\ref{fig:platform}\textcolor{blue}{c}, and corresponds to a design comprising 28 Mach-Zehnder interferometers (MZIs) with an overall set of 56 thermo-optic phase shifters \cite{ceccarelli2020low}. Each MZI represents a fully reprogrammable unit cell (see inset in Fig.~\ref{fig:platform}\textcolor{blue}{c}), where the two phase shifts can be used to program an arbitrary SU(2) transformation between the optical modes \cite{Clements2016}. 
The internal structure of the 12-mode chip utilizes a more compact arrangement of phase shifters within the MZI meshes \cite{Bell_optimal}, while ensuring full universality and reconfigurability thanks to 132 thermal heaters. In Fig.~\ref{fig:platform}\textcolor{blue}{d}, we present for the 12-mode device a simplified internal structure equivalent to that of the 8-mode device for ease of reference, as it provides a more intuitive illustration of the encoding scheme used for the spin simulator components.

\begin{figure*}[t]
    \centering
    \includegraphics[width=\linewidth]{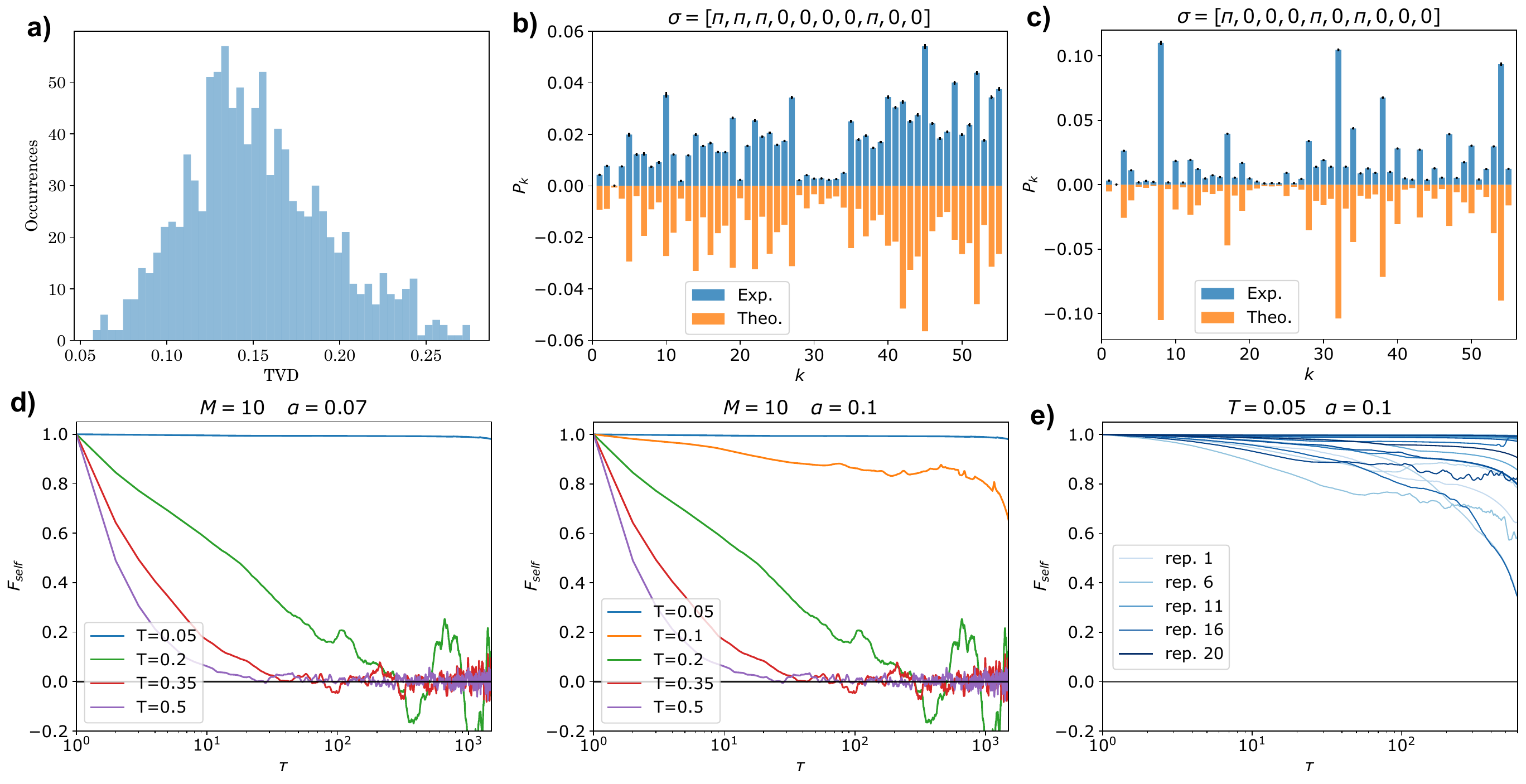}
    \caption{\textbf{Experimental characterization of the 10-spin simulator within the 12-mode integrated device.} a) Histograms of the 1024 TVD values between the theoretical and the two-photon experimental distributions for each spin configuration and for a fixed $10\times10$ scattering matrix $S$. b-c) Details on the experimental two-photon distributions for two spin settings. TVD is $0.139 \pm0.002$ in b) and $0.057\pm0.002$ in c). The uncertainties derive from the Poissonian statistics of the single-photon counts.  d) Spin-spin autocorrelation function $F_{\rm self}(\tau)=\langle \bm\sigma(t)\cdot\bm\sigma(t+\tau)\rangle$ for various temperatures of the MC simulation and different choices of the parameter $\alpha=N_P/10^2$. We selected $N_P=7$  ($\alpha = 0.07$)  
    and  $N_P=10$ 
    ($\alpha = 0.1$) output configurations.  
    e) Details on the $20$ replicated Metropolis dynamics with the same $S$ matrix, at $T=0.05$ and $\alpha=0.1$ . Each replica starts from a random initial spin configuration.
    } 
    \label{fig:12mode_results}    
\end{figure*}

\paragraph*{\textbf{Universal 8-mode device: 6 spins simulator.}}

The architecture proposed in Ref.~\cite{zanfardino2025} for a photonic simulation of a $M$-spin system envisages a first transformation that creates coherent and unbiased superpositions in the space associated with {$n_{\mathrm{ph}}$} photons distributed over 
{$M$} modes, when the photons are initially prepared in a single mode. The transfer matrix that satisfies such a requirement is the Discrete Fourier Transform (DFT), whose elements are defined as follows {$D_{jl}=M^{-1/2} e^{-2\imath\pi jl/M}$}. The $M$-mode DFT can be, in general, encoded in a $M$-port linear interferometer according to the universal decompositions of Refs.~\cite{Clements2016, Bell_optimal} via $M$ layers of concatenated MZIs. 
The requirement of an unbiased splitting over the $M$ modes can be satisfied also by a simplified circuit that envisages only $M/2$ layers of MZIs. This circuit can split the signal entering from a fixed input port over the $M$ modes with equal amplitude as required by the proposal of Ref.~\cite{zanfardino2025}, thus reproducing the first column of the DFT matrix. Such a simplified layout allows us to encode the full circuit for the simulation of a $M=6$ spin system into the 8-mode device, as reported in Fig.~\ref{fig:platform}\textcolor{blue}{b}. 
Then, we allocate 6 phase shifters (blue squares in Fig.~\ref{fig:platform}\textcolor{blue}{b}) for the spin encoding, followed by the random evolution encoded in the remaining layers of the device. 

The investigation started by evaluating the agreement between the expected two-photon distributions and the ones recorded in the experiment. Given the small size of the problem, it is still possible to measure all the possible configurations of the spin layer. 
In Fig.~\ref{fig:8mode_results}, we report 3 examples of the 64 spin configurations and the associated two-photon probability distribution for a fixed random $S$ scattering matrix. 
In Fig.~\ref{fig:8mode_results}\textcolor{blue}{b} we report the experimental Metropolis MC dynamics at temperatures $T=0.1$, $0.25$, $0.35$, $0.5$ (cfr. Fig.~\ref{fig:concept}\textcolor{blue}{b}), and the behavior of the time autocorrelation function of the spins $F_{\rm self}(\tau)=\langle \bm\sigma(t)\cdot\bm\sigma(t+\tau)\rangle$ with the number of Monte Carlo steps. A Monte Carlo step (MCS) consists in {$M$ Monte Carlo updates, that is $M$ spin-flip proposals \cite{Newman1999}. The $\tau$ time unit corresponds to one MCS, while the energy $H_S(\bm \sigma|\mathcal{K})$ is given by Eq.~\eqref{p_Hamiltonian}.
The memory storage capacity $\alpha$ is defined as the ratio between the size $N_P=|\mathcal K|$ of the set $\mathcal{K}$ and 
{$M^{n_{\mathrm{ph}}}$}, $\alpha = |\mathcal{K}|/M^2$ with $n_{ph}=2$. Fig.~\ref{fig:8mode_results}\textcolor{blue}{b} shows the MC runs for different values of $T$ and $\alpha$, with $N_P=$ 1, 3, 6.


\paragraph*{\textbf{Universal 12-mode device: 10 spins simulator.}}
An analogous experimental setup is employed to simulate a spin system of higher dimension. To this end, a 12-mode universal device is programmed as illustrated in Fig.~\ref{fig:platform}\textcolor{blue}{d}. The first part splits  the signal homogeneously over $10$ modes, then $10$ phase shifters encode the spin system, and the last part of the circuit represents the random evolution $S$. Such an architecture can simulate $M=10$ spins. 
In Fig.~\ref{fig:12mode_results} we report the experimental results of a two-photon experiment carried out for a fixed $10 \times 10$ scattering matrix  $S$. We are still in the scenario in which it is possible to measure all the two-photon distributions corresponding to the 1024 10-spin settings (see Fig.~\ref{fig:12mode_results}\textcolor{blue}{a-c}). In panel d) we report the trends of the $F_{\mathrm{self}}(\tau)$ for various temperature and $\alpha$ values. The energy at each Metropolis MCS is estimated by selecting a set of two-photon output configurations. In particular, we consider firstly $N_P=7$ configurations ($\alpha = 0.07$) and then $N_P=10$ ($\alpha=0.1$). For this last $\alpha$ value we report also the behavior of $20$ replicas of the Metropolis dynamics performed with the same $S$ matrix at $T=0.05$, but starting from different initial random spin configurations.

 Taking advantage of the 12-mode device larger size, we use the 10-spin simulator to conduct a detailed study of various 
 {collective} phases. In particular, we aim at observing memory retrieval effects typical of Hopfield neural networks, as we show in the next section.

\begin{figure*}[t]
    \centering
    \includegraphics[width = 0.98\linewidth]{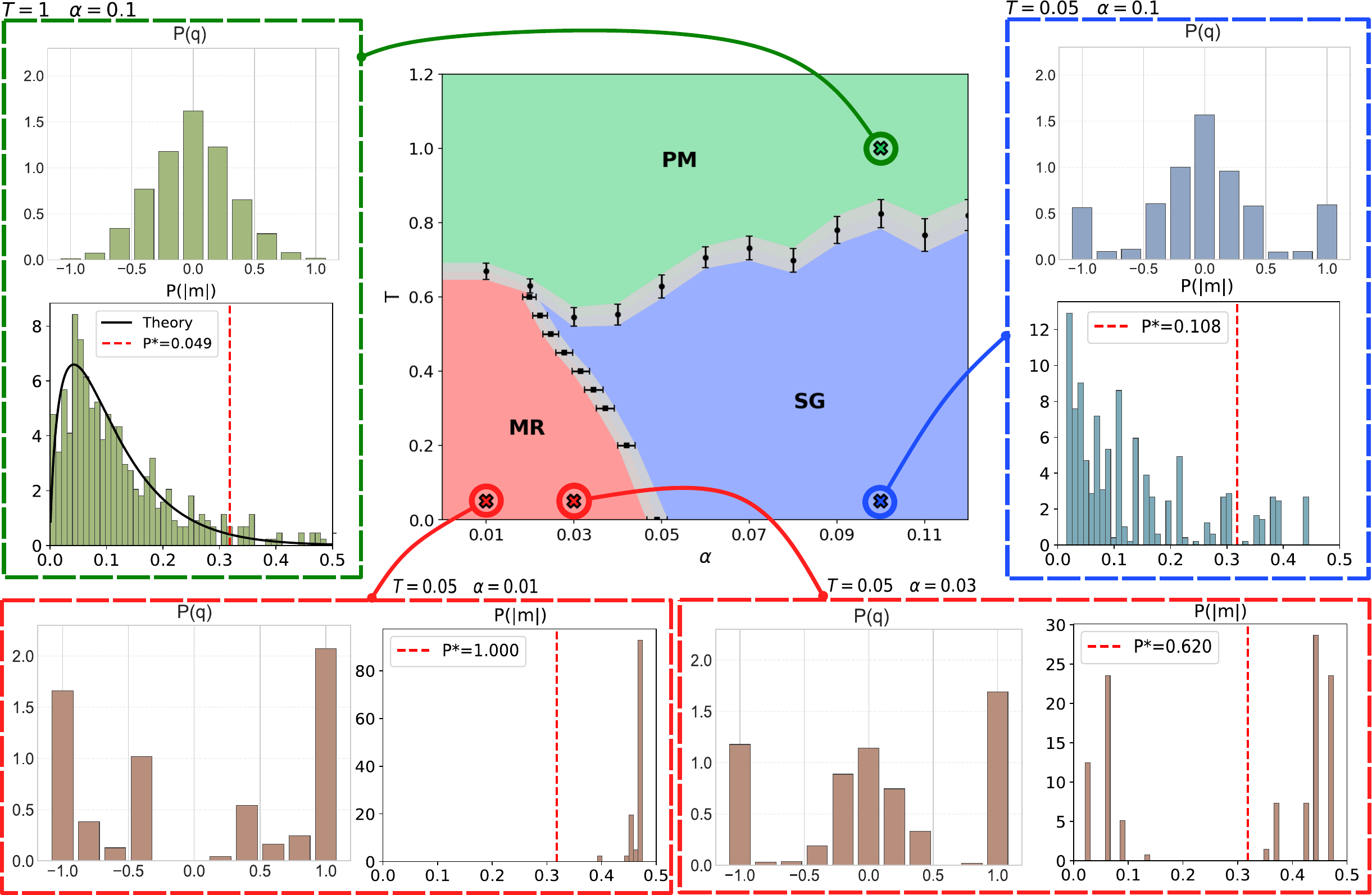}
    \caption{\textbf{Phase Diagram of the 10-spin simulator.} The $\alpha$ vs $T$ phase diagram was reconstructed from 360 simulated points obtained over $100$ different disorder realizations, corresponding to distinct measured scattering matrices $S$. Specifically, for each point $(\alpha,T)$ and for each realization of $S$, $N_{\mathrm{rep}}=100$ independent replicas were generated to reconstruct the overlap distributions $P(q)$ and the magnetization distribution $P(|m|)$, from which the memory retrieval (MR, red), spin-glass (SG, blue), and paramagnetic (PM, green) phases were identified. 
   In the $P(|m|)$ plots the vertical line is $|m|_{N_P}=1/\pi$ (see text). The phase boundaries were determined by identifying the transition between phases along the temperature and storage-capacity directions, with memory retrieval identified by a non-trivial enhancement of $P(|m|)$ with respect to the infinite-temperature reference and paramagnetic behavior detected from the emergence of Gaussian features in $P(q)$. For each boundary point, the critical temperature or storage capacity was averaged over the different realizations of disorder, while the shaded gray regions and error bars denote the corresponding standard error of the mean. 
    The circled crosses highlight four representatives of the experimentally measured points ($T=1$ and $\alpha=0.1$, $T=0.05$ and $\alpha = 0.01, 0.03, 0.1$) at fixed $S$ matrix. 
    The statistics of the overlap distribution $P(q)$ and the magnetization distribution $P(|m|)$ were reconstructed from 50 replicas of experimental Metropolis runs. 
    The displayed MR/SG transition curve is intended as a finite size proxy for a first order phase transition line. 
    Spinodals are not shown but one can observe from the $P(q)$, $P(|m|)$ that the $(0.03,0.05)$ point is a retrieved phase coexisting with a SG phase: $P(|m|)$ is non-zero  for $|m|>|m|_{N_P}$ and $P(q)$ also displays a related peak around $q=0$. For the PM phase, we reported the comparison between the experimental histogram and the expected $P(|m|)$ distribution.
    }
    \label{fig:phase_diag}
\end{figure*}

\section*{Observation of associative-memory retrieval phase}

The concept of memory retrieval is at the core of a Hopfield neural network (HNN). The network recognizes a pattern, that is, {\it retrieves a memory}, if it outputs such a pattern when starting from a different input configuration. If this happens not just for one particular input state, but for many different input states, 
then the memorized pattern  is an attractor in the configuration space with some finite basin of attraction. 
In a classical HNN, one is able to plant the memory patterns into the system; similarly, in the photonic spin simulator, we can implant a number of memories equal to $N_P$. In both cases, one can define a storage capacity $\alpha$ of the network, i.e. a properly rescaled number of patterns we are asking the system to store. As new attractors are added to the Hamiltonian of the system, that is, as the capacity grows, the energy landscape changes and new undesired local minima may appear, acting as spurious attractors for the evolution of the system. The basins of attraction of spurious patterns may become so large or so numerous that the system is effectively unable to reconstruct the memorized patterns consistently. In this case, we say that the network is behaving as a spin glass (SG) and is in a SG phase. 

In the present photonic HNN, the network dynamics is governed by the same interplay between memory attractors and spurious minima discussed above. For this specific architecture, the largest number of implantable memory patterns scales as the cardinality of the set of all photonic configurations $\mathcal C$, $\sim M^{n_{\mathrm{ph}}}$. Therefore, we have defined the storage capacity as $ \alpha= N_P/M^{n_{\mathrm{ph}}}$ \cite{zanfardino2025}.

Alongside the capacity $\alpha$, the energy landscape is also probed by introducing synaptic noise in the network, in the form of a temperature $T$ in the Metropolis evolution. 
At any finite $T$, the system may escape from the aforementioned local minimal configurations (whether they are the desired memorized patterns or the spurious ones that characterize the SG state) and the probability of this occurrence increases with $T$. At a high enough $T$, the system will display a paramagnetic (PM) behavior, where the network output is going to look completely random. 
 
In order to differentiate between the three behaviors, memory retrieval (MR), spin glass (SG) and paramagnetic (PM) phases, we create $N_\mathrm{rep}$ copies of our system, refereed to as ``replicas", each initialized with a random neuronal activity $\bm \sigma^{(a)}$, with $a=1,\dots,N_{\mathrm{rep}}$. We then perform Metropolis optimization independently on each replica (see Fig.~\ref{fig:12mode_results}\textcolor{blue}{e}). To discriminate the different regimes we introduce two  parameters: the bilinear magnetization {$m_{\vec{k}}$ of pattern $\mathbb X^{\vec k}$}, see Eq.~\eqref{X}, 
 \begin{eqnarray}
{m_{\vec k}} &=& {\sum_{\vec{x}} \mathbb X^{(\vec k)}_{\vec x}
\prod_{i=1}^{n_{\mathrm{ph}}}\sigma_{x_i} = \sum_{x_1\leq x_2}^{1,M} \sigma_{x_1} \mathbb X^{(k_1,k_2)}_{x_1,x_2} \sigma_{x_2}} \, ,
\label{M}
\end{eqnarray}
and the overlap $q_{ab}$ between replicas $a$ and $b$,
\begin{eqnarray}
\label{q}
q_{ab} &\equiv& \frac{1}{M} \sum_i \sigma_i^{(a)} \sigma_i^{(b)}\, .
\end{eqnarray}
The memory overlap $m_{\vec k}$ is a complex-valued quantity and is invariant under $\bm\sigma\to -\bm\sigma$ symmetry. In the following we consider the absolute magnetization $|m_{\vec{k}}|$, whose distribution $P(|m|)$ will be the order parameter to identify memory retrieval in a given instance of the photonic HNN. The  parameter $q_{ab}$ quantifies the overlap, as measured by the scalar product, between two different final spin configuration replicas $(a,b)$ at equilibrium, i.e., two Metropolis simulations with the same 
quenched disorder induced by the $S$ matrix in the quantum photonic simulator. The order parameter of the spin-glass (SG) phase is its probability distribution $P(q)$. Both distributions  are needed to characterize the collective behavior of the system in the $(\alpha,T)$ phase diagram, see Fig.~\ref{fig:phase_diag}.

In the high $T$ paramagnetic (PM) phase, $P(|m|)$ is expected to be approximately exponential for large $|m|$ \footnote{The average distribution over unitary $S$ realizations is $P(|m|) =\frac{2}{M+1}e^{-|m| M} \frac{M-1}{M+1} 2 |m| M^2 K_0(\sqrt{2} |m| M)$, where $K_0$ is the modified Bessel function of the second kind of degree $0$.}. At very large $T$, in a system with $M$ spins and $N_P$ planted patterns, there will be a probability $N_P/2^{M-1}$ of randomly selecting a $\bm\sigma$ configurations optimally aligned to a $\mathbb X$ pattern in $\mathcal K$. Operatively, for each unitary $S$, we can compute the value $|m|_{N_P}$ corresponding to (at least) the $1-N_P/2^{M-1}$ quantile of $P(|m|)$. This will be the reference to discriminate a possible memory recovery at low $T$. 

While in the PM phase the distribution $P(q)$ is expected to be trivially peaked around $q=0$, in the SG phase it will exhibit a highly nontrivial behavior, featuring many peaks both at high and low values of $q$. This characteristic reflects the existence of many states in the configuration landscape of the Hopfield Hamiltonian Eq.~\eqref{p_Hamiltonian} and the possibility for different replicas to dynamically relax to the same state with maximum overlap $q$ or to different states of varying correlation, including those with essentially no overlap, $q\simeq 0$. None of these states, however, is associated to a planted attractor; therefore the distribution $P(|m|)$ will not be qualitatively different from the one of the PM phase.

In the memory retrieval (MR) phase, the distribution $P(|m|)$ will be mostly peaked on a few large values of $|m|$, those corresponding to the magnetization Eq.~\eqref{M} of the planted $\mathbb X$ patterns. The distribution $P(q)$ will reflect the propensity of the $\bm \sigma$ states to be aligned to the $\mathbb X$ attractors, thus featuring significant peaks at high values of $|q|$ peaks and a strongly suppressed peak at $q=0$.


In the central diagram of Fig.~\ref{fig:phase_diag}, we report a numerical simulation of the phase diagram associated with the 10-spin photonic simulator presented in this work. We computed the distributions of the order parameters $(q,m)$ on a grid of $360$ point values of the controlling parameters $(\alpha,T)$. The distributions $P(q)$ and $P(|m|)$ at each point of the diagram were averaged over $100$ replicas for $100$ different randomly extracted $S$ matrices. In the next step, we experimentally investigated the four points highlighted in Fig.~\ref{fig:phase_diag}. For these points, we report the distributions $P(q)$ and $P(|m|)$ averaged over replicas and one choice of the $S$ matrices achievable within the circuit shown in Fig.~\ref{fig:platform}\textcolor{blue}{d}. The experimental findings confirm that at low $T$ and low $\alpha$, the system is the in MR phase. 

Regarding the identification of the MR phase, a clarification is in order: while in the standard Hopfield model at $T=0$, successful retrieval yields a (linear) magnetization $m = \pm 1$ for any perfectly stored pattern, in our generalized photonic realization the expectation of the absolute value $|m|$ of the complex bilinear magnetization Eq.~\eqref{M} of the best alignment of $\bm \sigma$ will be significantly smaller \cite{leuzzi2022quantitative}.
Therefore, in order to identify the MR phase from the distribution of {$|m|$} one has to adopt a direct operational procedure.
After estimating a reasonable threshold value of $|m|_{N_P}$ from the high $T$ distribution (e.g., the $1-N_P/2^{M-1}$ quantile; though, more cautiously, we used a smaller and fixed value $|m|=1/\pi\simeq 0.318$ \footnote{In the best possible case, in which the real and imaginary parts of the entries of a $\mathbb X^{(\vec k)}$ have the same sign and all the elements of $\vec k$ are equal (bunched), its expected value is $\mathbb E\left[\underset{ \bm \sigma}{\max}\left|m^{(\vec k)}\right|\right]=1/\pi=0.3183$.}), a point is classified in the MR phase if the probability $P(|m|>|m|_{N_P}) \geq 1/N_P$ (meaning that at least one pattern is retrieved). As $P(|m|>|m|_{N_P}) \rightarrow 1$, one has retrieval of all planted memories (full MR phase). Otherwise, only a partial retrieval of patterns occurs and the system is in the coexistence region between MR and SG. 
At high $T$, the system will be in the paramagnetic phase: a $(\alpha,T)$ point is classified in the PM phase if $P(|m|)$ is compatible with the high temperature reference and the overlap distribution $P(q)$ is essentially a Gaussian centered in zero \footnote{{Operationally, the system is in the paramagnetic phase as long as the Kurtosis $\langle (q-\langle q\rangle)^4\rangle/ \langle (q-\langle q\rangle)^2\rangle^2 > 2.65$}}. Finally, the system will enter the SG phase, i.e. low $T$ and large enough $\alpha$, as $P(q)$ begins to develop tails and eventually side peaks, uncorrelated with the memory attractors, so that no retrieval can occur and $P(|m|)$ still displays a paramagnetic-like shape. 

As a strong agreement was observed in all cases between experiments and numerical simulations, we were able to successfully identify the three distinct regimes: a memory retrieval phase, a spin-glass memory “black-out” phase, and a paramagnetic phase.

To provide a more immediate insight into the feasibility of stronger memory retrieval, we also engineered the scattering matrix $S$ to emulate the standard Hopfield model, where memory patterns are vectors of Ising-like variables and memory retrieval is characterized by a magnetization of $\pm 1$. This might be accomplished by employing real-valued matrices whose elements are $\pm 1$ up to an overall normalization factor, namely, Hadamard matrices. Since, however, Hadamard matrices are strictly defined for dimensions one, two, and multiples of four, they cannot be directly applied to our $10$-spin system. However, considering the available circuit depth and its reconfigurability, we were able to implement a scattering matrix featuring two rows that satisfy the aforementioned properties.
Within this configuration, we have successfully stored at least two $\mathbb X^{\vec k}$ memories analogous to those of a standard Hopfield model, as illustrated in Fig.~\ref{fig:hopfield}. In this case the magnetization $|m_{k,k}|$ of a $\bm \sigma$ configuration aligned with a bunched pattern $\mathbb X^{(k,k)}$ built by a $S_{\vec{k}|\vec {x}}$ row with Ising-like entries, see Eq.~\eqref{X}, tends to $1$ in modulus as $T\to 0$, providing a much stronger signal.  

\begin{figure}[t]
    \centering
    \includegraphics[width=1\linewidth]{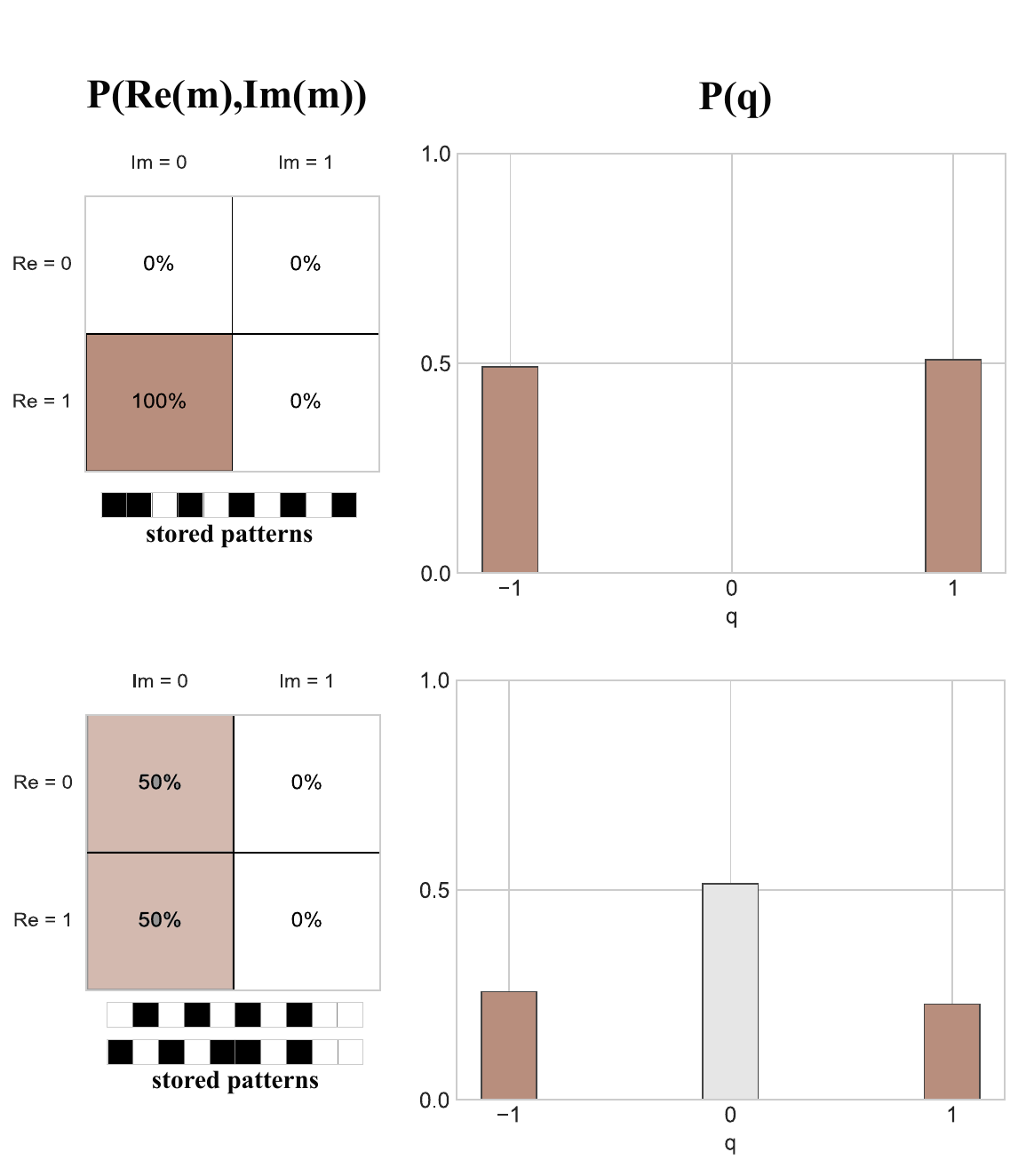}
    \caption{\textbf{Emulation of memory retrieval effects of the classical Hopfield model.} Experimental results for a scattering matrix featuring two rows with entries $\pm 1$ up to an overall normalization factors. The photonic spin-simulator displays the typical magnetization and order parameter $q$ behavior expected for a 4-body Hopfield model with respectively 1 stored pattern (above) and 2 orthogonal stored patterns (below).
    The table on the left showcases the joint probability of  Im(m) and Re(m): in both cases $m$ is real and only takes values $m=\pm1$ and $m=0$.}
    \label{fig:hopfield}
\end{figure}

\section*{Discussion and outlook}
This work presents the first experimental demonstration of associative memory retrieval using a photonic spin quantum simulator based on the generalized Hopfield model. Our photonic platform successfully reproduced the theoretical predictions for a four-body Hopfield network, clearly identifying three distinct operational phases: the memory retrieval phase at low storage capacities and temperatures, the spin-glass ``black-out" phase characterized by frustrated dynamics, and the paramagnetic phase where thermal fluctuations dominate. 

The experimental validation of memory retrieval demonstrates that quantum photonic systems can effectively simulate complex neural network dynamics that are computationally challenging on classical hardware for large systems. The key achievement lies in leveraging two-photon evolution within a programmable array of binary phase shifters to emulate Hopfield-like neuronal interactions. Our results confirm that at optimal operating conditions, the system consistently converges to fixed points with high memory overlap, successfully reconstructing stored patterns and demonstrating the practical viability of quantum simulation approaches for studying associative memory dynamics.

The demonstrated platform opens several promising research directions that could significantly advance both quantum simulation and neuromorphic computing. Immediate extensions {with the present devices} include: (i) scaling to higher-order Hopfield models 
{($p = 2 n_{\rm ph}> 4$), including large $p$}
to explore more effective memory storage mechanisms, up to exponential storage capacities
\cite{Demircigil2017,ramsauer2021,Lucibello2024};
(ii) use all available phase values of the device unit cell to build a vector Potts-Hopfield model in which $\sigma=e^{\imath \phi}$ are planar spins with up to $2^8$ directions, mimicking continuous $XY$ spins; (iii) implement and study mixtures of dense Hopfield models with different $p$; (iv) probe spin networks with lower connectivity.

Architectural developments will focus on implementing larger photonic circuits to accommodate significantly more spins. 
The integration of local and sparse connectivity patterns represents a natural evolution toward more realistic neural network topologies, moving beyond the current all-to-all coupling scheme.
Furthermore, using a number of modes  fit  to implement Hadamard unitary  matrices would allow to study Ising-spin-like memory patterns boosting both the retrieval signal and the storage capability.

Methodological advances may incorporate adaptive learning protocols and real-time pattern storage capabilities, transforming the platform from a static simulator to a dynamic quantum neural network. Long-term applications encompass, quantum machine learning algorithms, solving of optimization problems, and the development of fault-tolerant quantum neural networks. 
In conclusion, the present work establishes photonic quantum simulation as a viable pathway for understanding complex many-body dynamics in physical and neural systems, at the crossroad between quantum physics, statistical mechanics, and cognitive science.

\section*{Acknowledgments}

This work was supported by the PNRR MUR CN ICSC – Centro Nazionale di Ricerca in High Performance Computing, Big Data and Quantum Computing – and the PNRR MUR PE NQSTI – National Quantum Science and Technology Institute – both funded by the European Union – NextGenerationEU; and by the European Union Horizon Europe research and innovation program under the EPIQUE Project – Grant Agreement No. 101135288.  
S.P. acknowledges funding from VILLUM FONDEN (MapQP, No. VIL60743), the European Research Council (ERC StG ASPEQT, No. 101221875), Danmarks Innovationsfond research grant No. 4356-00009B (HyperTenQ), and the NNF Quantum Computing Programme. Fabrication of the laser-written processors was partially performed at PoliFAB, the micro and nano-fabrication facility of the Politecnico di Milano (http://www.polifab.polimi.it/). A.R., F.C. and R.O. would like to thank the PoliFAB staff for the valuable technical support.


%

\end{document}